\documentclass[aps,pre,twocolumn,showpacs,superscriptaddress,groupedaddress]{revtex4-2}
\pdfoutput=1
\usepackage{hyperref}
\usepackage{epsfig}
\usepackage{amsmath}
\usepackage{graphicx}
\usepackage{wrapfig}
\usepackage{dcolumn}
\usepackage{bm}
\usepackage{amssymb}
\usepackage{titlesec}
\usepackage{mathtools}
\usepackage{relsize}
\usepackage{cleveref}
\usepackage{calligra}
\usepackage{lipsum}

\usepackage[usenames,dvipsnames]{color}

\begin{document}

\title{Redirecting counter-moving swarms through collision}
\author{Jason Hindes$^{1}$, Chinthan B. Prasad$^{1}$, Loy McGuire$^{1}$, and Ira B. Schwartz$^{1}$}
\affiliation{$^{1}$U.S. Naval Research Laboratory, Washington, DC 20375, USA}
\email{jason.m.hindes.civ@us.navy.mil}
\begin{abstract}
Multi-swarm systems, where two or more swarms of mobile agents occupy the same region of space with different parameters and goals, occur in a variety of biological, engineering, and defense applications. Composites of multiple swarms can produce hybrid spatiotemporal patterns, which compared to single swarming systems, are relatively unexplored. In this work, we develop a framework for studying the collision of counter-moving swarms, each with its own preferred, stable velocity before collision. We show that redirection of such swarms after collision occurs when a stable velocity synchronized state of the multi-swarm composite exists. Using a rigid-body approximation, we are able to extract how scatter-redirection transitions scale with swarm parameters in a variety of scenarios from reciprocal and non-reciprocal systems to symmetric and antagonistic parameter values. Our results compare well to simulations of both particle modeled agents and wheeled robots. 
\end{abstract}
\maketitle

\section{\label{sec:Intro}INTRODUCTION}
Swarms consist of large numbers of mobile agents that interact in space to produce collective-motion patterns \cite{VICSEK201271,ActiveMatter,4200853,doi:10.1073/pnas.1005766107}. These systems can be designed, trained, or evolve in such a way that their collective spatiotemporal behaviors effectively solve a group-level problem\cite{10.1093/oso/9780195131581.001.0001,Yang2018,10.1371/journal.pone.0122827,Huang2025}. Remarkably, swarming dynamics is found on a very wide-range of space and time scales, and across many scientific and engineering disciplines, from cellular migration, foraging, and collective assembly\cite{Copeland2009BacterialSA,StatisticalPhysicsOfBacteriaSwarms,doi:10.1126/science.abj3065}, to swarming insects\cite{Giardina2008-wc,EquationOfStateForInsectSwarms}, flocking birds\cite{Ballerini2007InteractionRA,10.1371/journal.pcbi.1002894}, schooling fish\cite{10.1371/journal.pcbi.1002915,Calovi_2014}, and swarms of mobile robots for exploration, mapping, disaster recovery, and defense missions\cite{8722887,DroneSwarmStrategyForDetectionAndTracking,doi:10.1126/scirobotics.abd8668,9432146}.

An important problem for swarming research is to analyze and control a swarm's emergent pattern formation with certain target properties, without tightly engineering all aspects of the underlying agents\cite{Heuthe2025,wpgk-hhp7,mazesolvingdensitydrivenswarms,10610540}. Such swarming systems are often robust to failure, heterogeneity, uncertainty, time delays, and state-space perturbations\cite{10.1371/journal.pcbi.1002894,7063970,7403243,Hindes2021,Islam2023,PhysRevE.101.042202,doi:10.1073/pnas.2219948120,mousa2020}. In the manner of physical systems, one would like to control the pattern-formation of swarms in a low-dimensional or low-rank way, ideally for instance, by toggling a few control knobs or accessible parameters\cite{8901095,PhysRevE.109.014312}. Building such swarms of autonomous mobile agents-- somewhere between tightly engineered and black-box systems--
can benefit from an analytical approach.  

In a variety of applications, multiple swarms can occupy the same space, resulting in combined pattern-formation, scattering, capturing, redirection, etc. \cite{s22134773,9683410,PhysRevE.103.062602,Lama2025,drones6100271}. In the case of multiple colliding swarms, it has been shown that the relative velocities, numbers of agents, and coupling parameters between the swarms can be tuned to produce different collision outcomes\cite{ARMBRUSTER201745,PhysRevE.103.062602,10.1063/5.0159522,Kolon2018}. 
More generally, multi-species swarms are known to reshape the dynamical modes of single swarms and produce multi-stability of known collective behaviors\cite{PhysRevE.107.024607,gkhv-rp16}. Yet, much remains unknown about how combined patterns form in multi-swarm systems, and how pattern stability and bifurcations depend on the interactions within and between different swarms.

Motivated by these issues, we construct a modeling framework where two counter-moving swarms collide, each with different preferred velocities, and focus on the case where the two swarms combine and redirect their motion after collision. We formulate and analyze redirection in terms of the appearance of a stable velocity synchronized state for the combined swarm in parameter space. Using a rigid-body approximation, we show that the velocity synchronized state corresponds to a local minimum of an effective potential function, which depends on the spatial configuration of the swarm composite and the relative velocities of the two colliding swarms. This analysis reveals a variety of scaling patterns for different swarming systems. For example, in the case of head-on collision of swarms that differ only in their number of agents and velocities, we find that the minimum number of agents in one swarm needed to reverse the motion of a counter-moving swarm is independent of the latter's size. Moreover, for colliding swarms with antagonistic and reactive behavior, we find that redirection requires that a normalized measure of attraction between the swarms should be larger than an equivalent measure of repulsion-- independent of internal forces and spatial configurations before collision. These findings are shown to be in good agreement with particle and wheeled-robot simulations.

Our paper is organized as follows. In Sec.\ref{sec:PhysicsOverview} we introduce a model for the collision of counter-moving swarms with preferred velocities. In Sec.\ref{sec:VelocitySynch} we show how stable redirection of the colliding swarms emerges with a velocity synchronized state of the combined-swarm system, and develop a bifurcation theory based on rigid-body approximations. In Sec.\ref{sec:Examples} we explore the consequences of our bifurcation theory in several cases and test predictions with particle and robotic swarm simulations. In Sec.\ref{sec:Conclusion} we provide a discussion of our results and thoughts on next-steps and generalizations.


\section{\label{sec:PhysicsOverview} SWARMS WITH PREFERRED VELOCITIES}
The swarms of interest in this work will consist of simple mobile agents tasked with three basic objectives: to maintain group cohesion in space, to avoid contact that is too-close with others, and to move collectively at a certain velocity. Instead of being tightly engineered, these objectives are encoded in parameterized interactions among the agents and simple controls with local rules that produce behavior with the intended properties. 

Specifically, let us consider a swarm of $N$ agents in which every agent is accelerated by several state-dependent control forces and/or physical interactions, including self-propulsion, attraction, and repulsion. We assume that the self-propulsion force tends to select a preferred velocity for every agent, analogous to a natural frequency in a limit-cycle oscillator system\cite{RevModPhys.77.137,Kuramoto1984ChemicalOW}. Importantly, the preferred velocities may not be identical for all swarming agents in general. In addition, attraction and repulsion interactions express the tendency for agents to cluster in space with certain agents, while avoiding contact that is too-close with others. If we denote the position of the $i$th agent's center of mass in two spatial dimensions at time $t$, $\boldsymbol{r}_{i}(t)\!=\!(x_{i}(t),y_{i}(t))^{\top}$ and its velocity and acceleration $\dot{\boldsymbol{r}}_{i}$ and $\ddot{\boldsymbol{r}}_{i}$, respectively, a relatively simple model that captures the intended dynamics is
\begin{subequations}
\label{eq:BasicPhysics}
\begin{align}
&\ddot{\boldsymbol{r}}_{i}=\alpha_{i}[\boldsymbol{u}_{i}-\dot{\boldsymbol{r}}_{i}]+\sum_{j\neq i}\bold{F}_{ij}(\bold{r}_{j}\!-\!\bold{r}_{i}),\;\;\;\text{with}\\
&\bold{F}_{ij}(\bold{r}_{j}\!-\!\bold{r}_{i})\!=\!\frac{\boldsymbol{r}_{j}\!-\!\boldsymbol{r}_{i}}{|\boldsymbol{r}_{j}\!-\!\boldsymbol{r}_{i}|}\!\Bigg(\dfrac{a_{ij}}{l_{ij}^{(a)}}e^{-|\boldsymbol{r}_{j}-\boldsymbol{r}_{i}|/l_{ij}^{(a)}}\!\!\!-\!\dfrac{b_{ij}}{l_{ij}^{(b)}}e^{-|\boldsymbol{r}_{j}-\boldsymbol{r}_{i}|/l_{ij}^{(b)}}\!\!\Bigg)
\end{align}
\end{subequations}
where $\alpha_{i}$ is a characteristic self-propulsion constant for agent $i$, $\boldsymbol{u}_{i}$ is a preferred velocity for agent $i$\footnote{Note that each agent will approach its preferred velocity asymptotically when the attraction and repulsion interactions vanish}, $a_{ij}$ and $b_{ij}$ are attractive and repulsion constants, respectively, for agent $i$ with respect to agent $j$, and $l_{ij}^{(a)}$ and $l_{ij}^{(b)}$ are characteristic length scales for attraction and repulsion interactions for agent $i$ with respect to agent $j$ \cite{PhysRevE.63.017101,PhysRevLett.96.104302,Hindes_2024,1605401,Hindes2021,robotics12030081,Melchiorre2022RobotCA}. We note that the attraction and repulsion interactions are tangent to the displacement vector between two agents, and have magnitudes that are exponentially decaying functions of the distance between agents. This latter property encodes the tendency for local range-sensing to be maximal near a given agent, but fall-off rapidly over a characteristic detection range or zone. Other smooth, short-range functions are expected to produce similar dynamics.  

In a swarm robotics context, the preferred velocity may correspond to $\boldsymbol{u}_{i}=\hat{\boldsymbol{D}}_{i}/\tau_{i}$, for example, where $\hat{\boldsymbol{D}}_{i}$ is an estimated displacement with respect to a far-away target and $\tau_{i}$ is a preferred arrival time. Similar linear self-propulsion terms have been considered in engineering contexts\cite{1657384,doi:10.2514/1.25629,Montanari2025}, in contrast to nonlinear speed-selection mechanisms for undirected bio-physical systems\cite{PhysRevE.63.017101,PhysRevLett.96.104302}. On the other hand, typically, $l^{(a)}$ corresponds roughly to the sensing range of the robots, while $l^{(b)}$ is a collision avoidance length scale that is on the order of the robot's size\cite{4209425,10.1063/1.5142849,robotics12030081,Melchiorre2022RobotCA}. We point out that if the position-dependent interactions satisfy Newton's third law, then $a_{ij}\!=\!a_{ji}$ and $b_{ij}\!=\!b_{ji}$. We consider this {\it reciprocal} limit as a starting point, but consider non-reciprocal swarms as well, 
$a_{ij}\!\neq\!a_{ji}$ and $b_{ij}\!\neq\!b_{ji}$. Non-reciprocal and asymmetric interactions in swarms are part of a general class for active-matter systems\cite{doi:10.1073/pnas.2010318117,Lama2025}, 
which for instance, can arise if agent $i$ tries to avoid agent $j$, while agent $j$ tends to follow agent $i$, as in herding or predator-prey dynamics\cite{PhysRevLett.91.218102,PhysRevResearch.6.L032012}.

For single swarms with homogeneous parameters the basic  behaviors for systems such as Eqs.(\ref{eq:BasicPhysics}) are well-understood\cite{CHUANG200733,4209425,doi:10.1137/13091779X}. However, much less is known about multi-swarm dynamics 
when two or more swarms interact in the same region of space with potentially different parameters and goals. In such situations, it is possible for the swarms to combine and form new collective states, and for one swarm to redirect and/or capture another\cite{ARMBRUSTER201745,PhysRevE.103.062602,10.1063/5.0159522}. It is to these phenomena that we turn to in this work.     

\subsection{\label{sec:Setup} Collision of two flocking swarms}
In particular, we are interested in the collision of two swarms with agents behaving according to Eqs.(\ref{eq:BasicPhysics}). Initially, the two swarms are well-separated in space.
Moreover, within each swarm, the parameters in Eqs.(\ref{eq:BasicPhysics}) are taken to be homogeneous, but in general, the control constants, preferred velocities, numbers of agents, etc., may be different between the two swarms. 
For convenience, let us distinguish between the two swarms with color labels ``red" and ``blue" or $\text{R}$ and $\text{B}$, respectively. Namely, the red swarm consists of $N_{\text{R}}$ agents $\text{R}\!=\!\{1,2,...,N_{\text{R}}\}$, while the blue swarm consists of agents $\text{B}\!=\!\{N_{\text{R}}\!+\!1,N_{\text{R}}\!+\!2,...,N\}$. Agents in the red swarm have preferred velocities $\bold{u}_{i}\!=\!\bold{u}_{\text{R}}$ for $i\!\in\!\text{R}$ and $\bold{u}_{i}\!=\!\bold{u}_{\text{B}}$ for agents in the blue swarm, $i\!\in\!\text{B}$.

In addition, the two swarms are assumed to collide near the origin at time $t\!=\!T$, where $T\!\gg\!1$. In general upon collision, the spatial offset between the swarms is given by a parameter $\boldsymbol{\delta}$, which is assumed to be near zero 
. 
In terms of the initial conditions, the swarms are configured into {\it two flocks}, such that $\bold{r}_{i}(t\!\ll\!T)\!=\!\bold{u}_{\text{R}}(t-T)+\bold{z}_{i}^{*}+\boldsymbol{\delta}$ for $i\!\in\!\text{R}$
and $\bold{r}_{i}(t\!\ll\!T)\!=\!\bold{u}_{\text{B}}(t-T)+\bold{z}_{i}^{*}$ for $i\!\in\!\text{B}$, where $\bold{z}_{i}^{*}$ is a stable solution of a flocking configuration with vanishing self-propulsion force in Eqs.(\ref{eq:BasicPhysics})\cite{4209425,CHUANG200733,ARMBRUSTER201745, PhysRevE.103.062602, doi:10.1137/100804504}. If we define an $N\!\times\!N$ 
binary matrix $I$ whose elements are $I_{ij}\!=\!1$ if agents $i$ and $j$ have the same color and zero otherwise, then $\bold{z}_{i}^{*}$ satisfies
\begin{align}
\label{eq:TwoFlockConfig}
\bold{0}_{i}=\!\!\sum_{j\neq i}\!I_{ij}\bold{F}_{ij}(\bold{z}_{j}^{*}\!-\!\bold{z}_{i}^{*}) \;\;\;\forall i,
\end{align}
with $\sum_{i\in\text{R}}\bold{z}_{i}^{*}\!=\!\sum_{i\in\text{B}}\bold{z}_{i}^{*}\!=\!\bold{0}$.

\begin{figure*}[!htbp]
\center{\includegraphics[scale=0.68]{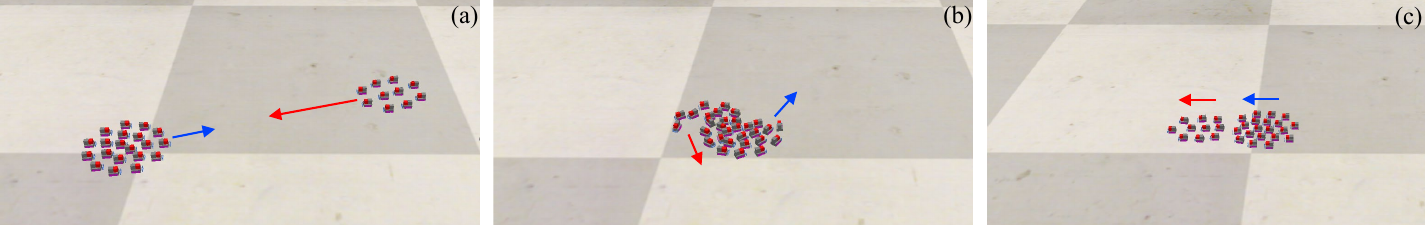}}
\caption{Snapshots of two swarms that are redirected upon collision: (a) before, (b) during, and (c) after. In (a), each swarm travels as a fixed-formation flock with a velocity equal to the preferred velocity of the respective swarm. In (b), the swarms collide and form a transient mill. In (c), the swarms form a stable composite that travels at a uniform velocity, which is different than the preferred velocities of either swarm. The agents are simulated, differential-drive robots with target-velocity controls given by Eqs.(\ref{eq:BasicPhysics})
. The arrows indicate representative velocities for agents within each swarm (blue and red). Note that in this case, the collision is head on.}
\label{fig1}
\end{figure*}

Figure \ref{fig1} illustrates the collision of two swarms with the setup described. In the first snapshot (a), a 10-agent red swarm approaches a 20-agent blue swarm from a configuration satisfying Eqs.(\ref{eq:TwoFlockConfig}). In the second snapshot (b), the swarms collide and form a transient milling pattern. In the third snapshot (c), the swarms coalesce into a single, stable flock that reverses the direction of the blue swarm. The agents are differential-drive robots (DDRs), which were simulated using CoppeliaSim software\cite{6696520} with target-velocity controls given by Eqs.(\ref{eq:BasicPhysics}); see Sec.\ref{sec:CoppeliaSim} for further details. 
Model parameters are $\alpha\!=\!4$, $a\!=\!0.1$, 
$b\!=\!0.1$, $l^{(a)}\!=\!2$, $l^{(b)}\!=\!0.1$, $\bold{u}_{\text{R}}\!=\!(-0.16,0)^{\top}$, and $\bold{u}_{\text{B}}\!=\!(0.05,0)^{\top}$. Note that in this example, all pairwise constants are the same $\forall (i,j)$. 
Throughout this work, the physical units for the DDR simulations are MKS. 

\section{\label{sec:VelocitySynch} STABLE REDIRECTION AS VELOCITY SYNCHRONIZATION}
Our primary goal is to examine the outcomes of two-swarm collisions described in Sec.\ref{sec:Setup}, and determine quantitative principles that allow us to predict and control the outcomes. For general parameters, the two swarms typically either {\it scatter} or {\it redirect} after collision. For scattering, the swarms collide but a composite state that stably contains all agents is not accessible, and so the red and blue swarms separate into  mutually uncoupled flocking configurations with altered trajectories. On the other hand, for redirection, such a stable composite is found, and the swarms merge together into a different formation that moves at a new collective velocity. Our main focus will be to determine the parameter values, such as the preferred velocities and the number of agents in the swarms, for which redirection can be expected. 

\begin{figure*}[!htbp]
\center{\includegraphics[scale=0.9]{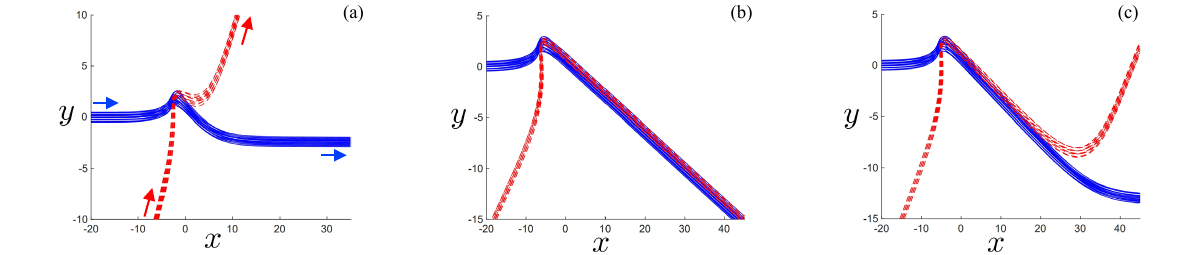}}
\caption{Outcomes for two-swarm collisions with simulated, wheeled robots as the preferred velocity of the red swarm is varied. Shown are trajectories in the $(x,y)$ plane for agents within each swarm: red and blue. The outcomes are: (a) scattering after collision, (b) redirection with a stable composite, and (c) weak instability of a two-swarm composite. Arrows in panel (a) show the direction of motion as time evolves.}
\label{fig2}
\end{figure*}

In Fig.\ref{fig2}, we show trajectories from DDR simulations for representative collision examples with parameters that produce: (a) scattering, (b) redirection, and (c) a transition between the two behaviors. In the examples shown, agents have similar controls and interactions within their respective swarms, but opposite interactions for agents in the counter-moving swarm. Namely, red swarming agents are attracted to blue agents, while blue swarming agents are repelled by red agents. Parameters for Eqs.(\ref{eq:BasicPhysics})
are $\alpha\!=\!3$, $a_{\text{R}\text{R}}\!=a_{\text{B}\text{B}}\!=\!a_{\text{R}\text{B}}\!=\!b_{\text{R}\text{R}}\!=\!b_{\text{B}\text{B}}\!=\!b_{\text{B}\text{R}}\!=\!0.1$,
$a_{\text{B}\text{R}}\!=\!b_{\text{R}\text{B}}\!=\!0$, $l^{(a)}_{\text{R}\text{R}}\!=\!l^{(a)}_{\text{B}\text{B}}\!=\!2$,  
$l^{(b)}_{\text{R}\text{R}}\!=\!l^{(b)}_{\text{B}\text{B}}\!=\!0.2$, $l^{(a)}_{\text{R}\text{B}}\!=\!l^{(b)}_{\text{B}\text{R}}\!=\!3$,
and $\bold{u}_{\text{B}}\!=\!(0.1,0)^{\top}$. 
For each plot, we choose a different value for the preferred velocity of the red swarm $\bold{u}_{\text{R}}\!=\!(0.1,v)^{\top}$, where: (a)  $v\!=\!0.1$, (b) $v\!=\!0.05$, and (c) $v\!=\!0.06$.  

\subsection{\label{sec:GeneralFixedPoint} Composite  flocking state}
In order for all agents to be redirected after collision, a stable, composite flock containing all agents must exist for a given set of parameters. If no such state exists, then the two swarms will scatter after a transient period. Note that existence and stability of a composite are necessary but not sufficient conditions for stable redirection, since the initial conditions specified in Sec.\ref{sec:Setup} 
may not lie in a given composite's basin of attraction\cite{PhysRevE.103.062602}. 

To make progress in understanding redirection, we begin by describing composite flocking states. In general, such states correspond to configurations in which the displacements between {\it all agents} are fixed while agents travel together at a constant velocity. 
By substituting the ansatz $\bold{r}_{i}(t\!\gg\!T)\!=\!\bold{U}t+\bold{d}_{i}^{*}$ into Eqs.(\ref{eq:BasicPhysics}), where $\bold{d}_{i}^{*}$ specifies the time-independent spatial configuration with a collective velocity $\bold{U}$, and taking $\ddot{\bold{r}}_{i}\!=\!\bold{0}_{i}\;\forall i$, we find that 
\begin{subequations}
\label{eq:GeneralComposite}
\begin{align}
&\bold{U}=\frac{1}{N}\sum_{i}\!\Bigg[\bold{u}_{i}+\frac{1}{\alpha_{i}}\sum_{j\neq i}\bold{F}_{ij}(\bold{d}_{j}^{*}\!-\!\bold{d}_{i}^{*})\Bigg]\;\;\;\text{and}\\ 
&\bold{0}_{i}=\alpha_{i}[\boldsymbol{u}_{i}-\bold{U}]+\sum_{j\neq i}\bold{F}_{ij}(\bold{d}_{j}^{*}\!-\!\bold{d}_{i}^{*})\;\;\;\forall i.
\end{align}
\end{subequations}
Equations (\ref{eq:GeneralComposite}) give us a numerically solvable $2N$-dimensional system for the two-swarm composite by fixing the origin  $\sum_{i}\bold{d}_{i}^{*}\!=\!\bold{0}$. Notice that unlike the state described by Eqs.(\ref{eq:TwoFlockConfig}), the self-propulsion force acting on agents does not vanish since $\bold{u}_{i}\neq\bold{U}\; \forall i$. For reference, note that if the interactions are reciprocal with forces between agents that are equal and opposite, $\bold{F}_{ij}\!=\!-\bold{F}_{ji}$, and self-propulsion constants that are also equal, then the composite velocity is simply equal to the average preferred velocity, $\bold{U}\!=\!\sum_{i}\!\bold{u}_{i}/N$.

We point out the similarity between Eqs.(\ref{eq:GeneralComposite}) and synchronized states of Kuramoto phase-oscillator systems with heterogeneity in the preferred velocities playing the role of quenched disorder in natural frequency\cite{RevModPhys.77.137}. For sufficiently large coupling, Kuramoto networks form a state of global frequency synchronization, in which the oscillators rotate at a single collective frequency\cite{DORFLER20141539}.  
Drawing on this analogy, we say that in the colliding swarms case, stable redirection corresponds to a state of {\it velocity synchronization}.  

The next step in understanding redirection is to study the stability of velocity-synchronized states (VSSs) satisfying Eqs.(\ref{eq:GeneralComposite}). In particular, we can determine the fixed-point bifurcations of solutions to Eqs.(\ref{eq:GeneralComposite}) 
using a linear approximation. Namely, we substitute $\bold{r}_{i}(t)\!=\!\bold{U}t+\bold{d}_{i}^{*}+\boldsymbol{\epsilon}_{i}(t)$ into Eqs.(\ref{eq:BasicPhysics}), 
and collect terms to $\mathcal{O}(\boldsymbol{\epsilon})$:
\begin{align}
\label{eq:LinearStability}
&\ddot{\boldsymbol{\epsilon}}_{i}=-\alpha_{i}\dot{\boldsymbol{\epsilon}}_{i}+\sum_{j}J_{ij}\boldsymbol{\epsilon}_{j},\;\;\;\text{with}\nonumber \\
& J_{ij}= \begin{cases}\;\;\;\;\dfrac{\partial\bold{F}_{ij}}{\partial(\bold{d}_{j}^{*}\!-\!\bold{d}_{i}^{*})}, & \text{if $j\neq i$}.\\
    -\mathlarger{\sum_{k\neq i}}\dfrac{\partial\bold{F}_{ik}}{\partial(\bold{d}_{k}^{*}\!-\!\bold{d}_{i}^{*})}, & \text{otherwise}.
  \end{cases}
\end{align}
The spectrum $\boldsymbol{\epsilon}_{i}(t)\!=\!e^{\lambda t}\bold{v}_{i}$, where $\bold{v}_{i}$ is an eigenmode of Eqs.(\ref{eq:LinearStability}), gives us information about the growth, decay, and oscillations of perturbations to a VSS. The spectrum depends principally on the $2N\!\times\!2N$ 
matrix $J$. 

Here, we point out a few important properties of Eqs.(\ref{eq:LinearStability}). First, because of translational invariance, the matrix $J$ has at least two zero eigenvalues for all parameter values. Second, if the interactions are reciprocal, then $J$ is symmetric and has real eigenvalues only. In this case, the appearance of redirection corresponds to a saddle-node bifurcation\cite{10.5555/289919,Kuramoto1984ChemicalOW,Manik2014}. In particular, if we denote the ordered eigenvalues of $J$: $\sigma_{1}\!\geq\!\sigma_{2}\!\geq\!\sigma_{3}\!\geq\!...\!\geq\!\sigma_{2N}$, where $\sigma_{1}\!=\!\sigma_{2}\!=\!0$, then as we approach the saddle-node bifurcation for reciprocal swarms, \textcolor{black}{$\sigma_{3}\!\rightarrow\!0$}. More generally, fluctuations around a VSS in reciprocal swarms, $\bold{r}_{i}(t)\!=\!\bold{U}t\!+\!\bold{d}_{i}(t)$, are effectively governed by a gradient system, $\ddot{\bold{d}}_{i}=-\alpha_{i}\dot{\bold{d}}_{i}-\partial V^{(\text{rec})}\!/\partial \bold{d}_{i}$, with an effective potential function
\begin{align}
\label{eq:V}
V^{(\text{rec})}=\;\;\;&\frac{1}{2}\sum_{i,j\neq i}\!\Big(b_{ij}e^{-|\boldsymbol{d}_{j}-\boldsymbol{d}_{i}|/l_{ij}^{(b)}}\!\!\!-\!a_{ij}e^{-|\boldsymbol{d}_{j}-\boldsymbol{d}_{i}|/l_{ij}^{(a)}}\Big) \nonumber \\
&-\sum_{i}\alpha_{i}(\bold{u}_{i}\!-\!\bold{U})^{\!\top}\!\bold{d}_{i}\;, \end{align}
similar to many phase-oscillator networks\cite{RevModPhys.77.137,10.1063/1.5041377}. Hence, when interactions are reciprocal, VSSs correspond to configurations that are local minima of Eq.(\ref{eq:V}). On the other hand, for non-reciprocal interactions, the matrix $J$ is not symmetric and has complex eigenvalues in general. Thus, for non-reciprocal colliding swarms, a VSS can change stability, not only through saddle-node bifurcations, but Hopf bifurcations as well. 

In order to predict whether redirection can occur as swarm parameters are varied, therefore, one can  numerically solve Eqs.(\ref{eq:GeneralComposite}-\ref{eq:LinearStability}), and track when VSS eigenvalues cross the imaginary axis\cite{10.5555/289919,Kuramoto1984ChemicalOW}. However, this approach has downsides including: high-dimensionality for large swarms, the potential for multiple VSS solutions, and the possibility that any particular solution may not be accessible from the colliding-swarms initial conditions specified in Sec.\ref{sec:Setup}. To make further progress, we develop a simplifying approximation that allows us to find analytical relationships and scaling patterns for redirection.

\subsection{\label{sec:FixedFormation} Rigid-body approximation for redirection}
Near the transition separating scattering and redirection, we notice that the two swarms tend to settle onto a configuration (after transients) that resembles separated flocks, but with a particular spatial offset, $\boldsymbol{\Delta}^{\!*}$, that emerges after collision; see Fig.\ref{fig1} (c). Namely, 
\begin{align}
\label{eq:RGA}
& \bold{d}_{i}^{*}\approx \begin{cases}\bold{z}_{i}^{*}+\boldsymbol{\Delta}^{\!*}, & \text{if $i\in\text{R}$}.\\
    \bold{z}_{i}^{*} & \text{if $i\in\text{B}$}.
  \end{cases}
\end{align}
Compared to any general solution of Eqs.(\ref{eq:GeneralComposite}), the proposed Eq.(\ref{eq:RGA}) is clearly compatible with the colliding-swarms setup, and therefore we expect it to be potentially useful in capturing 
VSS behavior relevant for collision. We note that because the internal configurations for each swarm are taken to be approximately fixed during collision, the ansatz Eq.(\ref{eq:RGA}) is effectively a {\it rigid-body approximation} (RBA) for the scatter-redirection transition\cite{999971152002121,CHUANG200733}.

In order to fully specify the RBA, we need to determine the unknown parameter $\boldsymbol{\Delta}^{\!*}$.
In general, the RBA is not a formal solution of Eqs.(\ref{eq:GeneralComposite}). 
Nevertheless, we expect the RBA to apply on average, i.e, to the centers of mass for each swarm. To find $\boldsymbol{\Delta}^{\!*}$, therefore, we take the average of Eqs.(\ref{eq:GeneralComposite}) (b) over red and blue agents, respectively. The result is a (vector) fixed-point equation for each swarm's center of mass:
\begin{subequations}
\label{eq:RigidBodyTwoEquations}
\begin{align}
&\bold{0}=\alpha_{\text{R}}[\boldsymbol{u}_{\text{R}}-\bold{U}]+\frac{1}{N_{\text{R}}}\sum_{i\in\text{R}}\sum_{j\neq i}\bold{F}_{ij}(\bold{d}_{j}^{*}\!-\!\bold{d}_{i}^{*}),\;\;\;\text{and}\\ 
&\bold{0}=\alpha_{\text{B}}[\boldsymbol{u}_{\text{B}}-\bold{U}]+\frac{1}{N_{\text{B}}}\sum_{i\in\text{B}}\sum_{j\neq i}\bold{F}_{ij}(\bold{d}_{j}^{*}\!-\!\bold{d}_{i}^{*}).
\end{align}
\end{subequations}

Next, we point out that if the spatial configurations satisfy Eq.(\ref{eq:RGA}) and Eqs.(\ref{eq:TwoFlockConfig}), then there is a cancelation of all interactions between agents within the same swarm. This allows us to further simplify Eqs.(\ref{eq:RigidBodyTwoEquations}). Namely, let us subtract the blue swarm center-of-mass equation-- divided by the self-propulsion constant-- from the same equation for the red swarm. Doing so, we find that $\boldsymbol{\Delta}^{\!*}$ is the solution of a simple two-dimensional equation  
\begin{align}
\label{eq:RigidBodyCompact}
&\bold{0}=\bold{u}_{\text{R}}\!-\!\bold{u}_{\text{B}}+\frac{1}{\alpha_{R}}\left<\bold{F}_{\text{B}\rightarrow\text{R}}\right>(\boldsymbol{\Delta}^{\!*})
-\!\frac{1}{\alpha_{B}}\left<\bold{F}_{\text{R}\rightarrow\text{B}}\right>(\boldsymbol{\Delta}^{\!*}),\nonumber \\
&\text{where} \nonumber\\
&\left<\bold{F}_{\text{B}\rightarrow\text{R}}\right>(\boldsymbol{\Delta})=\frac{1}{N_{\text{R}}}\sum_{i\in\text{R}}\sum_{j\in\text{B}}\bold{F}_{ij}(\bold{z}_{j}^{*}\!-\!\bold{z}_{i}^{*}\!-\!\boldsymbol{\Delta})\;\;\; \text{and}\nonumber\\
&\left<\bold{F}_{\text{R}\rightarrow\text{B}}\right>(\boldsymbol{\Delta})=\frac{1}{N_{\text{B}}}\sum_{i\in\text{B}}\sum_{j\in\text{R}}\bold{F}_{ij}(\bold{z}_{j}^{*}\!+\!\boldsymbol{\Delta}\!-\!\bold{z}_{i}^{*}). 
\end{align}


Our final step for analyzing redirection in the general RBA is to determine the appropriate bifurcations. Redirection is expected when there is a stable solution for $\bold{\Delta}^{\!*}$ in Eq.(\ref{eq:RigidBodyCompact}). Similar to Sec.\ref{sec:GeneralFixedPoint}, we can make progress by analyzing the Jacobian of Eq.(\ref{eq:RigidBodyCompact}) with respect to variation in $\boldsymbol{\Delta}^{\!*}$. Compared to the general system, the $2\times2$ Jacobian matrix of Eq.(\ref{eq:RigidBodyCompact}) has only real eigenvalues since $\boldsymbol{F}_{ij}(\boldsymbol{z}_{j}^{*}\!-\!\boldsymbol{z}_{i}^{*}\!+\!\boldsymbol{\Delta})$ can be written as a gradient,  
\begin{align}
&\boldsymbol{F}_{ij}(\boldsymbol{z}_{j}^{*}\!-\!\boldsymbol{z}_{i}^{*}\!+\!\boldsymbol{\Delta})=\frac{\partial}{\partial\boldsymbol{\Delta}} \Big(\!-a_{ij}e^{-|\boldsymbol{z}_{j}^{*}\!-\boldsymbol{z}_{i}^{*}+\boldsymbol{\Delta}|/l_{ij}^{(a)}}\;+\nonumber \\
&b_{ij}e^{-|\boldsymbol{z}_{j}^{*}\!-\boldsymbol{z}_{i}^{*}+\boldsymbol{\Delta}|/l_{ij}^{(b)}}\Big).
\end{align}
The gradient property implies that the Jacobian 
is a Hessian matrix of second derivatives with respect to $\bold{\Delta}^{\!*}$ \cite{Binmore_Davies_2002}. Thus, the off-diagonal elements are equal, since the order of partial derivatives does not matter for smooth functions. As a consequence, 
redirection emerges in a saddle-node bifurcation when 
\begin{align}
\label{eq:RGAbifurcation}
\det\Bigg(\frac{1}{\alpha_{\text{R}}}\dfrac{\partial \left<\bold{F}_{\text{B}\rightarrow\text{R}}\right>}{\partial\boldsymbol{\Delta}}-\frac{1}{\alpha_{\text{B}}}\dfrac{\partial \left<\bold{F}_{\text{R}\rightarrow\text{B}}\right>}{\partial\boldsymbol{\Delta}}\!\Bigg)\bigg|_{\boldsymbol{\Delta}\!=\!\boldsymbol{\Delta}^{\!*,s}}\!\!=0,
\end{align}
where $\bold{\Delta}^{\!*,s}$ is the collision parameter at bifurcation\footnote{One can solve the nonlinear system Eq.(\ref{eq:RigidBodyCompact}) and Eq.(\ref{eq:RGAbifurcation}) numerically with, e.g., {\footnotesize MATLAB}'s fsolve function.}.



\section{\label{sec:Examples}EXAMPLE CASES AND COMPARISONS TO SIMULATIONS}
Equation (\ref{eq:RigidBodyCompact}) and Eq.(\ref{eq:RGAbifurcation}) give us a simple low-dimensional framework to test, predict, and control redirection in a variety of collision scenarios for two swarms, which we turn to in the following sections.

\subsection{\label{sec:Reversal} Reversal in collisions with symmetric interactions}
The first scenario that we explore in detail involves the collision of {\it symmetric} swarms, i.e, swarms with 
different numbers of agents and preferred velocities, but otherwise equal parameters. In addition, we first consider swarms moving in opposite directions along the same axis with head-on collision. In this case, redirection corresponds to one of the swarms effectively {\it reversing} direction after collision, as in the example Fig.\ref{fig1}.

\begin{figure}[h]
\center{\includegraphics[scale=0.239]{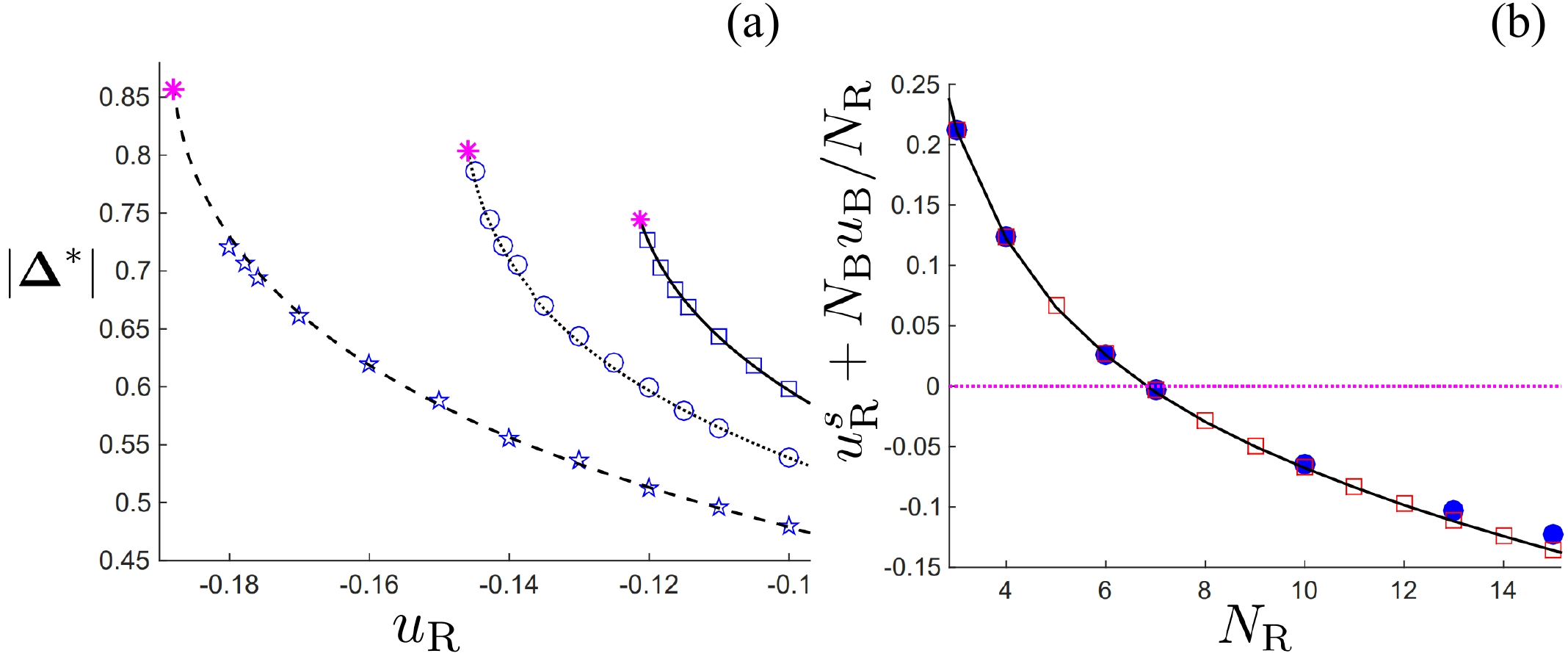}}
\caption{Bifurcation of symmetric swarms with head-on collision. 
(a) Time-averaged distance between the centers of mass for two swarms after collision vs. the red swarm's preferred velocity. Blue markers represent robot simulations (Sec.\ref{sec:CoppeliaSim}) with different numbers of red agents: $N_{\text{R}}\!=\!13$ (stars), $N_{\text{R}}\!=\!7$ (circles), and $N_{\text{R}}\!=\!3$ (diamonds); $N_{\text{B}}\!=\!20$ for each. Curves and magenta stars are predictions from Eq.(\ref{eq:RigidBodyCompact}) and Eq.(\ref{eq:RGAbifurcation}), respectively. 
(b) Bifurcation value of the red-swarm preferred velocity relative to the value required for reversal vs. the number of red agents. Robot and particle simulation results are shown with blue circles and red squares, respectively. Predictions  from solving Eq.(\ref{eq:RigidBodyCompact}) and Eq.(\ref{eq:RGAbifurcation}) are plotted with a solid curve. Other parameters are specified in Sec.\ref{sec:Reversal}.}
\label{fig3}
\end{figure}

Without loss of generality, 
we orient the preferred velocities for each swarm along the $x$-axis, i.e., $\boldsymbol{u}_{\text{R}}\!=\!(u_{\text{R}},0)^{\top}$ and $\boldsymbol{u}_{\text{B}}\!=\!(u_{\text{B}},0)^{\top}$. Similarly, the collision parameter aligns along the same axis,  $\boldsymbol{\Delta}^{\!*}\!=\!(\Delta^{\!*},0)^{\top}$. To start, we compare the solutions of the RBA Eq.(\ref{eq:RigidBodyCompact}) to DDR simulations. In Fig.\ref{fig3} (a) we plot the time-averaged value of the distance between the centers of mass for the two swarms after they collide versus the preferred velocity of the red swarm. Each set of blue plot markers corresponds to simulations with different numbers of red agents/robots: $N_{\text{R}}\!=\!13$ (stars), $N_{\text{R}}\!=\!7$ (circles), and $N_{\text{R}}\!=\!3$ (diamonds). For all examples $N_{\text{B}}\!=\!20$. Other model parameters are the same as those specified for Fig.\ref{fig1}. In addition, the solution of Eq.(\ref{eq:RigidBodyCompact}) is plotted with a curve for each example, along with the saddle-node bifurcation point plotted in magenta from solving Eq.(\ref{eq:RGAbifurcation}). In general, the agreement between theory and simulations is quite good, despite both the simplifications implicit in the RBA, and unmodeled dynamics of the robots that are not included in the target-velocity controller Eqs.(\ref{eq:BasicPhysics}) \cite{github_repo}. 

Qualitatively, we observe that the faster the red swarm moves in the opposite direction compared to the blue swarm, the larger the offset between their centers of mass in the VSS. Roughly speaking, as the separation between the two swarms approaches an inflection point of the positional interactions, such interactions are no longer sufficient to compensate for the different swarm velocities, and the saddle-node bifurcation occurs. 
The critical value for the separation between the swarms, $\bold{\Delta}^{\!*}\!\rightarrow\!\boldsymbol{\Delta}^{\!*,s}$, is reached, for example, when the velocity of the red swarm approaches a limit, $\boldsymbol{u}_{\text{R}}\!\rightarrow\!\boldsymbol{u}_{\text{R}}^{s}$, for which Eq.(\ref{eq:RigidBodyCompact}) and Eq.(\ref{eq:RGAbifurcation}) are satisfied simultaneously. 
We test these predictions in Fig.\ref{fig3} (b) by comparing $u_{\text{R}}^{s}$ to the largest, negative value of a red swarm's preferred velocity for which a VSS is observed in simulations. 
DDR and Eqs.(\ref{eq:BasicPhysics}) simulation results 
are shown with blue circles and red squares, respectively. Both are in good agreement with the RBA.


\subsubsection{Scaling with swarm parameters}
We note that in Fig.\ref{fig3} (b), $u_{\text{R}}^{s}$ is plotted relative to the value that produces a VSS with zero velocity. Namely, $U\!=\!0$ implies $u_{\text{R}}\!=\!-N_{\text{B}}u_{\text{B}}/N_{\text{R}}$. The $U\!=\!0$ line is plotted in magenta in Fig.\ref{fig3} (b) for reference. The general pattern is that as the number of agents in the red swarm, $N_{\text{R}}$, increases, the VSS reverses direction at a point where the black and magenta curves intersect. Note that when the VSS reverses direction, so does the blue swarm, which has a positive preferred velocity. As a general implication, if our goal is to parameterize a red swarm that reverses blue, for fixed attraction and repulsion parameters, there is some minimum number required, $N_{\text{R}}^{\text{(min)}}$, which in the example Fig.\ref{fig3} (b) corresponds to $N_{\text{R}}^{\text{(min)}}\!=\!7$.

An interesting question concerns how $N_{\text{R}}^{\text{(min)}}$
scales with swarm parameters. We can estimate the answer by considering the limit of large numbers $N_{\text{R}},N_{\text{B}}\!\gg\!1$. In this case, the continuum approximation becomes applicable, where the configuration of each swarm is described by the same density functional form in the RBA, $\rho(\bold{z})$ for a flock centered at $\bold{z}\!=\!\bold{0}$ (where we drop the subscript and $*$ notation for convenience). Note that for general models, the density functions for flocking-states with vanishing self-propulsion are analyzable using functional methods\cite{doi:10.1137/100804504,doi:10.1137/090749037,CHUANG200733}.  
By replacing the sum over particles in Eq.(\ref{eq:RigidBodyTwoEquations})(b) by integrals over density, taking $U\!=\!0$, and solving for $N_{\text{R}}$, we find the expected minimum number of red agents needed to reverse the blue swarm 
\begin{align}
\label{eq:MinimumNumber}
&N_{\text{R}}^{\text{(min)}}=\frac{\alpha u_{\text{B}}}{S(\Delta^{\!*,s})},\;\;\; \text{with}\nonumber\\
&S(\Delta)=-\!\int\!\!\!\int
\!F^{(x)}(\bold{z}'-\!\bold{z}\!+\!\Delta\hat{\bold{x}})
\rho(\bold{z})\rho(\bold{z}')d\bold{z}^{2}d\bold{z}'^{2}, 
\end{align}
where $\bold{z}$ and $\bold{z}'$ are displacement vectors from the blue and red swarm centers, respectively, $\hat{\bold{x}}$ is the unit vector $(1,0)^{\top}$, $\boldsymbol{\Delta}^{\!*,s}\!=\!\Delta^{\!*,s}\hat{\bold{x}}$ is the solution of Eq.(\ref{eq:RGAbifurcation}), 
$F^{(x)}(\bold{z}'-\!\bold{z}\!+\!\Delta\hat{\bold{x}})$ is the $x$-component of the force on a blue agent located at $\bold{z}$ from a red agent located at $\bold{z}'+\!\Delta\hat{\bold{x}}$, and $d\bold{z}^{2}$ and $d\bold{z}'^{2}$ are the differential area elements for the blue and red swarms, respectively. If we substitute the parameter values from Fig.\ref{fig3} (b) into Eq.(\ref{eq:MinimumNumber}), then we estimate the critical number 
to be $N_{\text{R}}^{\text{(min)}}\!\approx\!7.08$ \footnote{The integrals over density in this work were numerically approximated by simulating two separate flocking swarms, each with $1000$ agents, from random initial conditions with Eqs.(\ref{eq:BasicPhysics}) until $t\!=\!10^{4}$. Using each swarm's steady-state configuration, $\bold{z}_{\text{B}}$ and $\bold{z}_{\text{R}}$ for the blue and red swarms, respectively, the integral operator $\int\!\!\int\!\rho_{\text{B}}(\bold{z}_{\text{B}})\rho_{\text{R}}(\bold{z}_{\text{R}})d\bold{z}_{\text{B}}^{2}d\bold{z}_{\text{R}}^{2}$, can be estimated by $\frac{1}{N_{\text{B}}N_{\text{R}}}\sum_{i\in\text{B},j\in\text{R}}$, 
similar to a Monte Carlo integration approximation.}, which is very near the noted observational value. 

More important than the exact value for the denominator of Eq.(\ref{eq:MinimumNumber}) (which can be computed numerically), is the fact that the integral is a function only of the positional force parameters, $a,b,l^{(a)},$ and $l^{(b)}$. Specifically, $S(\Delta^{\!*,s})$ does not depend on the number of agents in the swarms or the preferred velocities. Hence, we learn that if the blue swarm doubles its velocity, $N_{\text{R}}^{\text{(min)}}$ also doubles. On the other hand, if the blue swarm doubles its number of agents, this has no effect on $N_{\text{R}}^{\text{(min)}}$. In fact, the red swarm does not need more agents for reversal; 
it needs agents that are twice as fast, according to $u_{\text{R}}\!=\!-N_{\text{B}}u_{\text{B}}/N_{\text{R}}^{\text{(min)}}$. 

\begin{figure}[t]
\center{\includegraphics[scale=0.239]{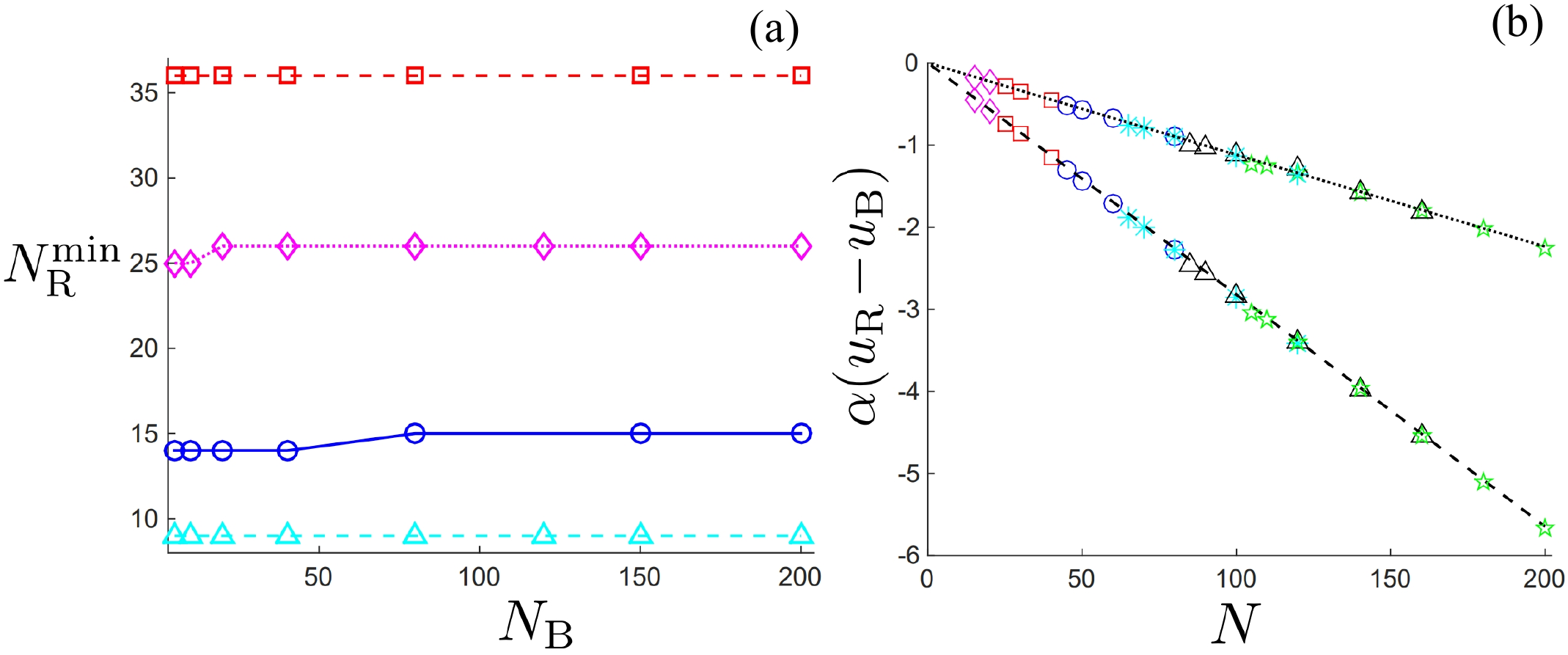}}
\caption{Scaling for reversal when two symmetric swarms collide for a wide range of model parameters. (a) Minimum number of red agents needed to reverse a blue swarm vs. the number of blue-swarm agents. Each plot label corresponds to a different set of ($\alpha$,$\;u_{\text{B}}$,$\;l^{(b)}$) parameters, detailed in Sec.\ref{sec:Reversal}. (b) Difference between the preferred velocities of red and blue swarms at bifurcation vs. the combined swarm size. Predictions from Eq.(\ref{eq:LinearInN}) are plotted with lines for each series. The top and bottom series correspond to
$l^{(b)}\!=\!0.1$ and $l^{(b)}\!=\!0.5$, respectively. Other parameters are specified in Sec.\ref{sec:Reversal}.}
\label{fig4}
\end{figure}

Several examples demonstrating these results are shown in Fig.(\ref{fig4}) (a). For four sets of parameters, we plot $N_{\text{R}}^{\text{(min)}}$
with different ($\alpha$,$\;u_{\text{B}}$,$\;l^{(b)}$):
blue circles ($1.0$,$0.4$,$0.1$), red squares ($1.0$,$0.5$,$0.5$), magenta diamonds ($1.3$,$0.6$,$0.1$), and cyan triangles ($0.5$,$0.2$,$0.5$). 
To find each plotted point, we fix $N_{\text{B}}$ and perform a series of collision simulations with Eqs.(\ref{eq:BasicPhysics}) and different $N_{\text{R}}\in\{1,2,...,50\}$ with $u_{\text{R}}\!=\!-N_{\text{B}}u_{\text{B}}/N_{\text{R}}\!-\!0.002$. The smallest value of $N_{\text{R}}$ that produces a stable VSS is plotted. Other model parameters are $a\!=\!b\!=0.1$ and $l^{(a)}\!=\!2$. The pattern in Fig.(\ref{fig4}) (a) follows our expectations based on Eq.(\ref{eq:MinimumNumber}). Namely, as $N_{\text{B}}$ grows large, there is effectively no change in $N_{\text{R}}^{\text{(min)}}$.

A similar analysis reveals that the critical preferred-velocity difference between the two swarms scales linearly with the total number of agents, $N\!=\!N_{\text{R}}+N_{\text{B}}\!\gg\!1$. In particular, by repeating the continuum approximation using Eq.(\ref{eq:RigidBodyCompact}), we find that 
\begin{align}
\label{eq:LinearInN}
\alpha(u_{\text{B}}\!-\!u_{\text{R}}^{s})=
N S(\Delta^{\!*,s})
\end{align}
with the same integral appearing as in Eq.(\ref{eq:MinimumNumber}). Thus, in the symmetric-reversal case, we learn that larger swarms are able to accommodate more discrepancy in the velocities of the composites.
These results are tested in Fig.\ref{fig4} (b) for two sets of repulsion length scales $l^{(b)}\!=\!0.1$ (upper) and $l^{(b)}\!=\!0.5$ (lower). Other force parameters are $a\!=\!0.1$, 
$b\!=\!0.1$, and $l^{(a)}\!=\!2$. For the plotted data, $\alpha$ ranges from $0.2$ to $2.0$, $u_{\text{B}}$ ranges from $0.2$ to $1.0$, $N_{\text{B}}$ ranges from $10$ to $100$, and $N_{\text{R}}$ ranges from $5$ to $100$. For each point, we fix all parameters except $u_{\text{R}}$, which is varied, and plot the most negative value for which a VSS is formed after collision in simulations of Eqs.(\ref{eq:BasicPhysics}). Equation (\ref{eq:LinearInN}) predictions are plotted with lines along side simulation results. We observe in Fig.\ref{fig3} (b) that the continuum approximation correctly predicts a scaling collapse of the data onto two lines with slopes given by the particular values of the integral in Eq.(\ref{eq:LinearInN}). 


\subsection{\label{sec:GeneralAngles}Redirection at general angles}
Next, we move on to consider collisions with more general angles, where the preferred swarm velocities are not on the same axis. Once again, our aim is to find appropriate red swarm parameters that redirect the blue swarm. Further scaling and analytical progress is possible when considering the limit in which the two colliding swarms have {\it isotropic mass distributions} in their flocking states before collision. Given the rotational symmetry of the attractive and repulsive forces in Eqs.(\ref{eq:BasicPhysics}), we 
observe isotropy when the swarm numbers are large $N_{\text{R}},N_{\text{B}}\!\gg\!1$ \cite{CHUANG200733}. 
In general, we note that the solution of the RBA Eq.(\ref{eq:RigidBodyCompact}) can be written as a 
local minimum of a potential function, $-\partial{V^{(\text{RBA})}}\!/\partial{\bold{\Delta}}|_{\bold{\Delta}=\bold{\Delta}^{\!*}}=\bold{0}$, where
\begin{align}
\label{eq:HIso}
V^{(\text{RBA})}(\bold{\Delta})=&-(\bm{u}_{\text{R}}\!-\!\bm{u}_{\text{B}})\!\cdot\!\bm{\Delta} \nonumber \\
&\!\!\!\!\!\!\!\!\!\!\!\!\!\!\!\!\!\!\!\!\!\!\!\!\!\!\!\!\!+\frac{1}{\alpha_{\text{R}}N_{R}}\!\sum_{\substack{i\in\text{R}, \\ j\in\text{B}}}\!\Big(b_{ij}e^{-|\bold{z}_{j}^{*}\!-\!\bold{z}_{i}^{*}\!-\boldsymbol{\Delta}|/l_{ij}^{(b)}}\!\!\!-\!a_{ij}e^{-|\bold{z}_{j}^{*}\!-\!\bold{z}_{i}^{*}\!-\boldsymbol{\Delta}|/l_{ij}^{(a)}}\Big) 
\nonumber \\
&\!\!\!\!\!\!\!\!\!\!\!\!\!\!\!\!\!\!\!\!\!\!\!\!\!\!\!\!\!+\frac{1}{\alpha_{\text{B}}N_{B}}\sum_{\substack{i\in\text{B}, \\ j\in\text{R}}}\!\Big(b_{ij}e^{-|\bold{z}_{j}^{*}\!-\!\bold{z}_{i}^{*}\!+\boldsymbol{\Delta}|/l_{ij}^{(b)}}\!\!\!-\!a_{ij}e^{-|\bold{z}_{j}^{*}\!-\!\bold{z}_{i}^{*}\!+\boldsymbol{\Delta}|/l_{ij}^{(a)}}\Big).
\end{align}
If the isotropic assumption is valid, then the summations in Eq.(\ref{eq:HIso}) do not depend on the direction of $\bold{\Delta}$. As such, the $(\bm{u}_{\text{R}}\!-\!\bm{u}_{\text{B}})\!\cdot\!\bm{\Delta}$ term 
is the only one that selects a spatial direction, and therefore, we expect the offset between the two swarms to align with their velocity difference, i.e., 
\begin{align}
\label{eq:DeltaIso}
\bold{\Delta}=\frac{\bold{u}_{\text{R}}-\bold{u}_{\text{B}}}{|\bold{u}_{\text{R}}-\bold{u}_{\text{B}}|}\;\Delta\equiv \hat{\bold{s}}\;\Delta.
\end{align} 

Given Eqs.(\ref{eq:HIso}-\ref{eq:DeltaIso}), the RBA bifurcation point Eq.(\ref{eq:RGAbifurcation}) in the isotropic approximation becomes the one-dimensional system 
\begin{align}
\label{eq:BifIso}
0=\frac{\partial^{2}V^{(\text{RBA})}}{\partial^{2}\Delta}\Big |_{\Delta=\Delta^{\!*,s}}\;\;.
\end{align}
Note that Eq.(\ref{eq:BifIso}) is satisfied for a particular critical length $\Delta^{\!*,s}$ (magnitude only), which is independent of the velocities of the swarms. 
Other patterns can be found by analyzing Eqs.(\ref{eq:HIso}-\ref{eq:BifIso}) in specific scenarios. 

\subsubsection{\label{sec:SymmetricGenAngles} Redirection angles for symmetric swarms}
As in Sec.\ref{sec:Reversal}, we begin with symmetric swarms. In particular, for symmetric and isotropic swarms with $N_{\text{R}},N_{\text{B}}\!\gg\!1$, the bifurcation condition Eq.(\ref{eq:BifIso}) is 
\begin{align}
\label{eq:BifIsoSymm}
0=\frac{\partial S}{\partial\Delta}\Big |_{\Delta=\Delta^{\!*,s}}\;\;.
\end{align}
Here, we use the fact that the interaction sums in Eq.(\ref{eq:HIso}) can be computed along any axis for $\bold{\Delta}$, i.e., $\hat{\bold{s}}=\hat{\bold{x}}$. 
Large-$N$ results allow us to calculate
the red-swarm parameters that deflect a blue swarm to a desired angle. To make parametric control with the red swarm one-dimensional, we assume that the red and blue swarms have orthogonal velocities, i.e, $\boldsymbol{u}_{\text{B}}\!=\!u_{\text{B}}\hat{\bold{x}}$ and $\boldsymbol{u}_{\text{R}}\!=\!u_{\text{R}}\hat{\bold{y}}$, where $\hat{\bold{y}}$ is the unit vector $(0,1)^{\top}$.  
In this case, the angle ($\phi^{*}$) at which the blue swarm is redirected after collision is $\phi^{*}\!=\!\arctan(\bold{U}\cdot\hat{\bold{y}}/\bold{U}\cdot\hat{\bold{x}})$.

In order for redirection at a given $\phi^{*}$ to be stable, $\Delta^{\!*}\!<\!\Delta^{\!*,s}$. For fixed blue-swarm parameters, $\Delta^{\!*}\!=\!\Delta^{\!*,s}$ corresponds to a maximum redirection angle $\phi^{*,s}$ that is achievable by the red swarm. The maximum angle can be found by combining the RBA solution in the isotropic and symmetric limits $-\partial{V^{(\text{RBA})}}\!/\partial{\Delta}|_{\Delta=\Delta^{\!*}}\!=\!0=|\boldsymbol{u}^{\text{R}}\!-\!\boldsymbol{u}^{\text{B}}|-NS(\Delta^{*})/\alpha\!=\!0$, with the bifurcation condition Eq.(\ref{eq:BifIsoSymm}). Solving for $\phi^{*,s}$ gives 
\begin{align}
\label{eq:MaximumAngle}
\phi^{*,s}=\arctan\Bigg(\frac{N_{\text{R}}}{N_{\text{B}}}\sqrt{\Bigg(\frac{NS(\Delta^{\!*,s})\!}{\alpha u_{\text{B}}}\Bigg)^{\!2}\!-\!1}\;\;\Bigg). 
\end{align}

\begin{figure}[t]
\center{\includegraphics[scale=0.230]{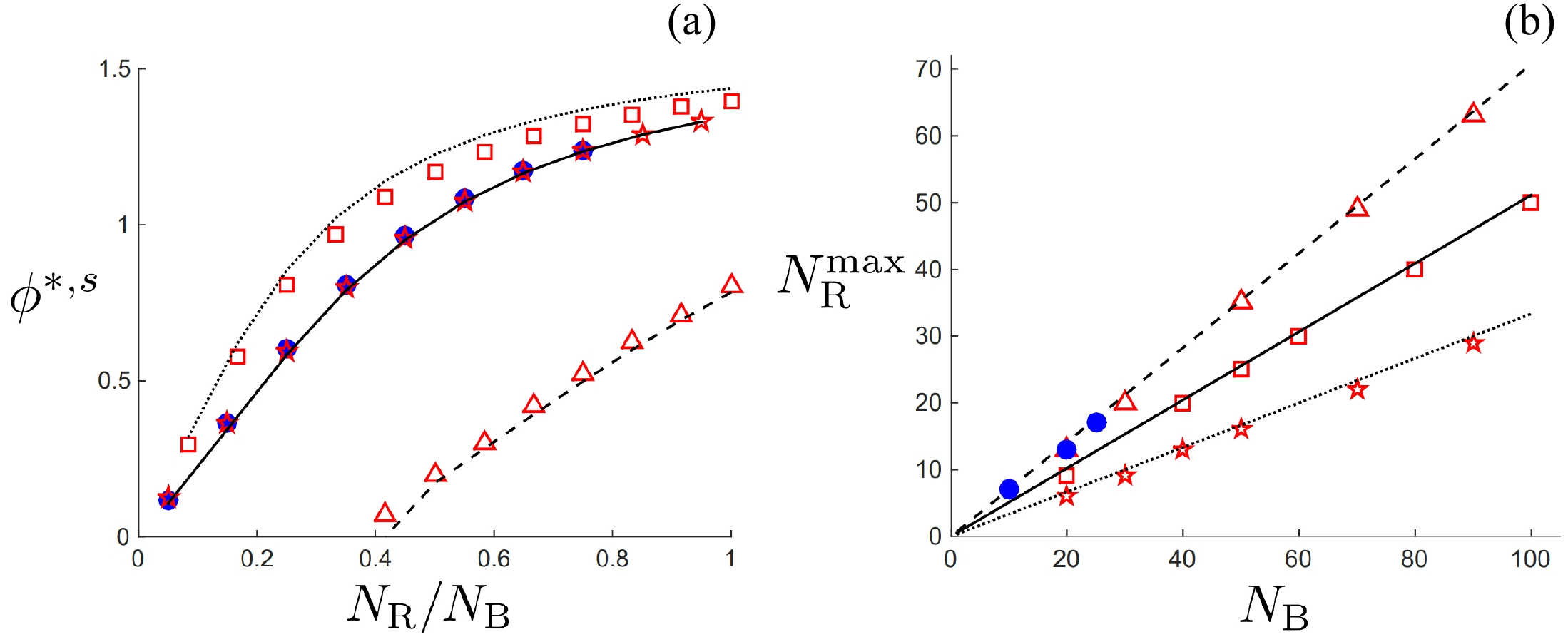}}
\caption{Redirection at general angles. 
Robot and Eqs.(\ref{eq:BasicPhysics}) simulations are plotted in blue and red, respectively. (a) Maximum redirection angle (radians) for symmetric swarms with orthogonal collision vs. the size of the red swarm relative to the blue. Each series corresponds to different $\alpha u_{\text{B}}$: $0.45$ (squares),  $0.25$ (circles and stars), and $2.4$ (triangles), as specified in Sec.\ref{sec:SymmetricGenAngles}. Curves represent Eq.(\ref{eq:MaximumAngle}) predictions. (b) Maximum number of red agents that can be embedded in a blue swarm for the antagonistic scenario. Each series corresponds to different ($\alpha_{\text{R}}$, $\alpha_{\text{B}}$, $a_{\text{RB}}$, $l^{(b)}_{\text{BB}}$): red triangles and blue circles ($4$, $3.3$, $0.06$, $0.1$), red squares ($4.5$, $3.5$, $0.046$, $0.2$), and red stars ($4.5$, $3$, $0.035$, $0.5$). Lines are predictions from Eq.(\ref{eq:AntagCond}). Other parameters are specified in Sec.\ref{sec:Antagonistic}. Note that for both (a) and (b), $\bold{u}_{\text{B}}\!=\!u_{\text{B}}\hat{\bold{x}}$.}
\label{fig5}
\end{figure}

If our goal is to parameterize a red swarm that redirects a blue swarm from its preferred trajectory, then Eq.(\ref{eq:MaximumAngle}) gives us an upper bound for the parametric control outcome. Figure \ref{fig5}(a) shows tests of Eq.(\ref{eq:MaximumAngle}), compared to DDR and Eqs.(\ref{eq:BasicPhysics}) simulations. Each series represents different values for $(\alpha,u_{\text{B}})$ pairs, from top to bottom, respectively: $(2.25,0.2)$, $(5,0.05)$, and $(8,0.3)$. Attractive and repulsive parameters are the same as in Fig.\ref{fig1}. For each series, we find the maximum value of $u_{\text{R}}$ that produces a VSS in simulations, and plot the corresponding angle at which the blue swarm is redirected. Note that the large-$N$ predictions from Eq.(\ref{eq:MaximumAngle}) are nearly identical to the RBA solutions of Eq.(\ref{eq:RigidBodyCompact}) and Eq.(\ref{eq:RGAbifurcation}), 
and so, we simply plot the former. 

Interestingly, the lower simulation dataset in Fig.\ref{fig5}(a) plotted with red triangles demonstrates that if 
blue's velocity controller, $\alpha u_{\text{B}}$, is too large, then there are certain composite swarms that are too small to stabilize redirection by a red swarm at any angle. In the noted example, $N_{\text{R}}/N_{\text{B}}\!\gtrsim\!0.4$. In fact, Eq.(\ref{eq:MaximumAngle}) tells us that 
\begin{align}
N_{\text{R}}+N_{\text{B}}>\frac{\alpha u_{\text{B}}}{S(\Delta^{\!*,s})},
\end{align} 
giving a number-bound for the controllable red swarm. Intuitively, the larger the red swarm in the symmetric scenario, the more effective the redirection of a blue swarm.

\subsubsection{\label{sec:Antagonistic} Antagonistic swarm collisions}
The final case that we consider involves colliding swarms where red agents are attracted to blue agents, while blue agents are repelled by red agents. Because the two swarms have opposing tendencies with respect to one another, we call this scenario {\it antagonistic}. 
In particular, we assume that red agents are attracted to blue agents with a coupling constant $a_{\text{RB}}$ and blue agents are repelled by red agents with a coupling constant $b_{\text{BR}}$, as in the example Fig.\ref{fig2}. Other inter-swarm couplings are taken to be zero $a_{\text{BR}}\!=\!b_{\text{RB}}\!=\!0$. 
In addition, to simplify the analysis, we make the further assumption that the sensing length scales and/or interaction zones for red and blue agents with respect to one another are similar in  the antagonistic scenario, i.e, $l_{\text{RB}}^{(a)}\!=\!l_{\text{BR}}^{(b)}\equiv\!l_{\text{RB}}$. 
Opposing interaction constants for the swarms with similar magnitudes can arise, for instance, if agents in one swarm effectively sense and react to changes in the motion of agents in the opposing swarm. 

Despite the opposing interactions, stable redirection can be observed as in the example Fig\ref{fig2}(b). For antagonistic redirection, the blue swarm is continuously repelled and deflected by the red swarm, which chases after it. Similar to the symmetric examples, we would like to parameterize the red swarm in order to produce such redirection. Here again, large-$N$ limits are useful for extracting scalings. Note that unlike symmetric scenarios, the two swarms are not assumed to have the same internal parameters, and therefore the density functions will not be the same in general. Let us denote the densities $\rho_{\text{R}}(\bold{z}_{\text{R}})$ for the red swarm and $\rho_{\text{B}}(\bold{z}_{\text{B}})$ for the blue swarm, given $N_{\text{R}},N_{\text{B}}\!\gg\!1$. By applying Eqs.(\ref{eq:HIso}-\ref{eq:DeltaIso}), and solving for $\Delta^{*}$ from $-\partial{V^{(\text{RBA})}}\!/\partial{\Delta}|_{\Delta=\Delta^{\!*}}\!=\!0$, we find that the collision parameter $\Delta^{*}$ for antagonistic swarms 
is the solution of
\begin{align}
\label{eq:RBA_iso_antag}
&\dfrac{|\bold{u}_{\text{R}}\!-\!\bold{u}_{\text{B}}|}{\dfrac{N_{\text{B}}a_{\text{RB}}}{\alpha_{\text{R}}}\!-\!\dfrac{N_{\text{R}}b_{\text{BR}}}{\alpha_{\text{B}}}}=A(\Delta^{*}), \;\;\;\;\text{where} \nonumber \\
&A(\Delta)=\!-\frac{\partial}{\partial\Delta}\!\int\!\!\!\int\!\!e^{-|\bold{z}_{\text{R}}-\bold{z}_{\text{B}}+\Delta\hat{\bold{s}}|/l_{\text{RB}}}\rho_{\text{B}}(\bold{z}_{\text{B}})\rho_{\text{R}}(\bold{z}_{\text{R}})d\bold{z}_{\text{B}}^{2}d\bold{z}_{\text{R}}^{2}. 
\end{align}


An important observation from Eq.(\ref{eq:RBA_iso_antag}) 
is that redirection cannot occur as 
the denominator goes to zero, since $A(\Delta)$ is finite. 
In this limit, only swarms with identical velocities $|\bold{u}_{\text{R}}\!-\!\bold{u}_{\text{B}}|\!=\!0$ can be stabilized. In fact, stability of a stable composite in the antagonistic scenario, generally, requires 
\begin{align}
\label{eq:AntagCond}
\frac{N_{\text{B}}a_{\text{RB}}}{\alpha_{\text{R}}}>\frac{N_{\text{R}}b_{\text{BR}}}{\alpha_{\text{B}}}. 
\end{align}
When Eq.(\ref{eq:AntagCond}) is not satisfied, the RBA Jacobian has at least one positive eigenvalue, making redirection unstable. Effectively, Eq.(\ref{eq:AntagCond}) expresses the intuitive requirement for redirection that the attractive forces acting on a typical red agent must be larger than the opposing repulsive forces acting on a typical blue agent (normalized by self-propulsion strength). 
Note that Eq.(\ref{eq:AntagCond}) holds generally and is independent of the internal parameters and preferred velocities of the red and blue swarms. 

In effect, Eq.(\ref{eq:AntagCond}) gives a
red swarm number-range for parametric control. 
Unlike the symmetric case, however, for antagonistic swarms, we find the opposite tendency: the red swarm cannot be too many in number in order to redirect the blue -- smaller is better. What is more, the maximum number of red swarming agents that can be embedded in the blue, $N_{\text{R}}^{(\text{max})}\!=\! (a_{\text{RB}}/b_{\text{BR}})(\alpha_{\text{B}}/\alpha_{\text{R}})N_{\text{B}}$, only depends on self-propulsion constants, inter-swarm coupling parameters, and the total number of blue agents. This result is tested in Fig.\ref{fig5}(b) against simulations of Eqs.(\ref{eq:BasicPhysics}) (plotted in red) and DDR simulations (plotted in blue). Each series represents a different combination of ($\alpha_{\text{R}}$, $\alpha_{\text{B}}$, $a_{\text{RB}}$, $l^{(b)}_{\text{BB}}$): red triangles and blue circles ($4$, $3.3$, $0.06$, $0.1$), red squares ($4.5$, $3.5$, $0.046$, $0.2$), and red stars ($4.5$, $3$, $0.035$, $0.5$). For each point, we increase $N_{\text{R}}$, while keeping all other parameters fixed, until a VSS is not observed after collision. Other model parameters are $a_{\text{R}\text{R}}\!=a_{\text{B}\text{B}}\!=\!b_{\text{R}\text{R}}\!=\!b_{\text{B}\text{B}}\!=\!0.1$, $b_{\text{B}\text{R}}\!=\!0.07$, $l^{(a)}_{\text{R}\text{R}}\!=\!l^{(a)}_{\text{B}\text{B}}\!=\!l_{\text{R}\text{B}}\!=\!2$, $l^{(b)}_{\text{R}\text{R}}\!=\!0.1$, $\bold{u}_{\text{B}}\!=\!(0.1,0)^{\top}$, and $\bold{u}_{\text{R}}\!=\!(0.1,0.002)^{\top}$. 

Altogether, we see that predicting the redirection of two colliding swarms with different preferred velocities is well-captured by stability analysis of a velocity synchronized composite -- even when the interactions between the swarms are heterogeneous and/or antagonistic.    

\section{\label{sec:Conclusion} DISCUSSION}
When two swarms interact in the same region of space, it is possible for one swarm to scatter, capture, redirect, or otherwise alter the motion of the other. In this work, we focused on the scenario in which two counter-moving and flocking swarms collide, each with their own preferred velocity as they approach the collision. For such swarms, we demonstrated how the existence of a stable velocity-synchronized composite of the combined swarms determines whether redirection can be expected after collision. By developing a rigid-body approximation, we were able to predict parameter values and scalings for the relative sizes, velocities, and coupling constants that separate scattering behavior from stable redirection in a variety of parameter regimes, including symmetric and antagonistic cases. For example, we were able to calculate scattering angles between symmetric colliding swarms using continuum approximations and the fact that velocity synchronized states within the rigid-body approximation can be described as minima of an effective potential function. Similarly, we found that there is an upper-bound for the number of agents that can be embedded into an antagonist  swarm, which depends only on inter-swarm couplings, self-propulsion constants, and the number of agents in the antagonist swarm. Our results were found to be in good agreement with both interacting-particle and wheeled-robot simulations.       

Our work represents progress on the general problem of predicting and controlling collision outcomes in multi-swarm systems. In addition, it is likely that the presented approach can shed light on more general swarming behavior, for example, involving generalized velocity synchronization in heterogeneous and networked swarms with linear velocity controls and quenched disorder. Yet, the results do not immediately apply to swarms with nonlinear self-propulsion, or to swarms with delay-dynamics from communication and control. Such systems can produce richer collective dynamics, including milling and chaotic behavior, and therefore the possibility of collision scenarios that involve two swarms with qualitatively different dynamics, e.g., flocks colliding with mills. Moreover, we studied non-reciprocal swarms with simple force-parameter heterogeneity and simple violation of Newton's third law with two-body interactions between agents. Extending the results to more general non-reciprocal dynamics\cite{doi:10.1073/pnas.2010318117,Lama2025} and collisions between swarms with more general collective behavior are interesting avenues for future work. Nevertheless, attacking such issues and implementing the results in robotics simulations and experiments can benefit from the work presented. 

\section*{ACKNOWLEDGEMENTS}
JH, IBS, CBP, and LM were supported by the U.S. Naval Research Laboratory funding \textcolor{black}{(N0001424WX00021)}. 
JH and IBS were also supported by the Office of Naval Research funding (N0001425GI01182) and (N0001425GI01158).\\




\begin{appendix}{$\;\;\;\;\;\;\;\;\;\;\;\;\;\;\;\;\;\;\;\;\;\;\;\;\;\;\;\;\;\;\;\;${\bf APPENDIX}}
\section{\label{sec:CoppeliaSim} CoppeliaSim Robot Simulations}
CoppeliaSim is a medium-fidelity simulation environment with built-in object collision handling, ideal for multi-agent applications\cite{6696520}. Individual agents consist of geometric 3D models and associated behavior scripts. The robots modeled in this work are differential-drive robots with a maximum angular velocity of 1 rad/s based on the Anki Vector mobile robotic platform. N x M agents can be initialized at a desired Euclidean spacing and initial heading. Velocity commands are generated every 50 ms by each agent using Eqs.(\ref{eq:BasicPhysics}). Each agent's state information is shared globally, enabling synchronized flocking behavior. As such, inter-agent communication is simplified in this study, but specific communication architectures can be implemented in subsequent studies. Simulation code associated with this work can be found at\cite{github_repo}.
\end{appendix}

\bibliography{RedirectingCollidingSwarms}

\end{document}